\documentclass[10pt,final,reqno]{amsart}
     \makeatletter
     \def\section{\@startsection{section}{1}%
     \z@{.7\linespacing\@plus\linespacing}{.5\linespacing}%
     {\bfseries
     \centering
     }}
     \def\@secnumfont{\bfseries}
     \makeatother
\newtheorem{theorem}{Theorem}[section]
\newtheorem{lemma}[theorem]{Lemma}
\newtheorem{proposition}[theorem]{Proposition}

\theoremstyle{definition}

\theoremstyle{remark}
\newtheorem{remark}[theorem]{Remark}
\numberwithin{equation}{section}
\usepackage[dvipdfmx]{graphicx}
\usepackage[dvipdfmx]{color}
\usepackage{amsfonts}
\usepackage{amssymb}
\usepackage{mathrsfs}
\usepackage{amsmath,amsthm}
\usepackage{amsmath}

\newcommand{\E}{\mathop {\rm E}}

\begin{document}

\title[Formulation of Optimal Execution with Uncertain Market Impact]{Mathematical Formulation of an Optimal Execution Problem with Uncertain Market Impact}

\author{Kensuke Ishitani}
\address{Kensuke Ishitani: Department of Mathematics, 
Faculty of Science and Technology, Meijo University, 
Tempaku, Nagoya 468-8502, Japan}
\email{kishitani@meijo-u.ac.jp}

\author[Takashi Kato]{Takashi Kato*}
\address{Takashi Kato: Division of Mathematical Science for Social Systems, 
Graduate School of Engineering Science, Osaka University, 
1-3 Machikaneyama-cho, Toyonaka, Osaka 560-8531, Japan}
\email{kato@sigmath.es.osaka-u.ac.jp}
\thanks{* This research was supported by a grant-in-aid from the Zengin Foundation for Studies on Economics and Finance.}

\subjclass[2010] {Primary 91G80; Secondary 93E20, 49L20.}

\keywords{Optimal execution, market impact, liquidity uncertainty, stochastic control, L\'evy process}

\begin{abstract}
We study an optimal execution problem with uncertain market impact
to derive a more realistic market model. 
We construct a discrete-time model as a value function for optimal execution. 
Market impact is formulated as the product of a deterministic part 
increasing with execution volume 
and a positive stochastic noise part. 
Then, we derive a continuous-time model as a limit of a discrete-time value function. 
We find that the continuous-time value function is characterized 
by a stochastic control problem with a L\'evy process. 
\end{abstract}

\maketitle

\section{Introduction}\label{sec_intro}

The optimal portfolio management problem is central in mathematical finance theory. 
There are various studies on this problem, and recently more realistic problems,  
such as liquidity problems, have attracted considerable attention. 
In this paper, we focus on market impact (MI), 
which is the effect of the investment behavior of traders on security prices. 
MI plays an important role in portfolio theory, and is also significant when we consider 
the case of an optimal execution problem, 
where a trader has a certain amount of security holdings (shares of a security held) 
and attempts to liquidate them before the time horizon. 
The optimal execution problem with MI has been studied in several papers 
(\cite {Alfonsi-Fruth-Schied,
Almgren-Chriss,Bertsimas-Lo,Gatheral,Gatheral-Schied,Subramanian-Jarrow} and references therein,) 
and in \cite{Kato} such a problem is formulated mathematically. 

It is often assumed that the MI function is deterministic. 
This assumption means that we can obtain information about MI in advance. 
However, in a real market it is difficult to 
capture the effects of MI without any estimation error. 
Moreover, it often happens that a high concentration of unexpected orders 
will result in overfluctuation of the price. 
The Flash Crash in the United States stock market is 
a notable example of unusual thinning liquidity: 
On May 6th, 2010, the Dow Jones Industrial Average plunged by
about 9\%, only to recover the losses within minutes. 
Considering the uncertainty in MI, it is thus more realistic 
and meaningful to construct a mathematical model of random MI. 
Moazeni et al. \cite {Moazeni-et-al} studied the uncertainty in MI 
caused by other institutions by compound Poisson processes, and 
then studied an optimization problem of expected proceeds of execution 
in a discrete-time setting. 
They considered the uncertainty in arrival times of large trades from other institutions; 
however, MI functions of decision makers themselves 
were given as deterministic linear functions so that 
the decision makers knew how their own execution affected the market price of the security 
(the coefficients of MI functions were regarded as 
``expected price depressions caused by trading assets at a unit rate''). 

In this paper, we generalize the framework in \cite{Kato}, 
particularly considering a random MI function. 
The model proposed in Section 2 in \cite{Kato} is derived as 
a limit of a discrete-time optimal execution problem. 
Specifically, as in Section A in \cite{Kato} 
we first define a discrete-time value function 
to explicitly describe the situation of each large-volume trade. 
Then, by taking the limit, we derive the continuous-time version of the value function, 
which is the main model of \cite{Kato}. 
In the present study, we introduce a noise term to a discrete-time MI function 
to investigate how the effect of uncertainty in the MI function appears in the continuous-time model 
as a time-scaling limit. We then find that 
the randomness of MI in the continuous-time model is described as a jump of a L\'evy process. 

The rest of this paper is organized as follows. 
In Section \ref {section_Model}, we present the mathematical formulation of our model. 
We set a discrete-time model of an optimal execution problem as our basic model and 
define the corresponding value function. 
We also give a convergence theorem of the value functions as our main result. 
Section \ref {sec_proof} contains all the proofs. 
We briefly conclude this paper in Section \ref {section_conclusion}.

\section{The Model and Main Result}\label{section_Model}

In this section, we present the details of the proposed model, 
which is based on the argument in Section A in \cite {Kato}. 
Let $(\Omega ,\mathcal {F}, P)$ be a complete probability space.  
$T > 0$ denotes a time horizon, and we assume $T = 1$ for brevity. 
We assume that the market consists of one risk-free asset (cash) and 
one risky asset (a security).
The price of cash is always $1$, which means that a risk-free rate is zero. 
The price of the security fluctuates according to a certain stochastic flow, 
and is influenced by sales performed by traders. 

First, we consider a discrete-time model with a time interval $1/n$. 
We consider a single trader who has an endowment of $\Phi _0>0$ shares of a security. 
This trader liquidates the shares $\Phi _0$ over a time interval $[0,1]$ considering the effects of MI with noise. 
We assume that the trader sells shares at only times $0, 1/n, \ldots , (n-1)/n$ for $n\in \Bbb {N} = \{1,2,3,\ldots \}$. 

For $l=0,\ldots ,n$, we denote by $S^n_l$ the price of the security at time $l/n$, 
and we also denote $X^n_l = \log S^n_l$. 
Let $s_0>0$ be an initial price (i.e., $S^n_0 = s_0$) and $X^n_0 = \log s_0$. 
If the trader sells an amount $\psi ^n_l$ at time $l/n$, the log price changes to 
$X^n_l-g^n_l(\psi ^n_l)$, and by this execution (selling) the trader obtains 
an amount of cash $\psi ^n_l S^n_l\exp (-g^n_l(\psi ^n_l))$ as proceeds. 
Here, the random function 
\begin{eqnarray*}
g^n_l(\psi , \omega ) = c^n_l(\omega )g_n(\psi ), 
\ \ \psi \in [0, \Phi _0], \ \omega \in \Omega 
\end{eqnarray*}
denotes MI with noise, which is given by the product of a positive random variable $c^n_l$ and 
a deterministic function $g_n : [0, \Phi _0]\longrightarrow [0, \infty )$. 
The function $g_n$ is assumed to be non-decreasing, continuously 
differentiable, and satisfying $g_n(0) = 0$. 
Moreover, we assume that $(c^n_l)_l$ is independent and identically distributed (i.i.d.), 
and therefore noise in MI is time-homogeneous. 
Note that if $c^n_l$ is a constant (i.e., $c^n_l\equiv c$ for some $c > 0$,) 
then this setting is the same as in \cite {Kato}. 

After trading at time $l/n$, $X^n_{l+1}$ and $S^n_{l+1}$ are given by 
\begin{eqnarray}\label{fluctuate_X}
X^n_{l+1}=Y\Big (\frac{l+1}{n} ; \frac{l}{n}, X^n_l-g^n_l(\psi ^n_l)\Big ), 
\ S^n_{l+1} = e^{X^n_{l+1}}, 
\end{eqnarray}
where $Y(t ; r,x)$ is the solution of the following stochastic differential equation (SDE) 
on the filtered space $(\Omega , \mathcal {F}, (\mathcal {F}^B_t)_t, P)$: 
\begin{eqnarray*}
\left\{
\begin{array}{l}
 	dY(t; r,x) = \sigma (Y(t; r,x))dB_t+b(Y(t; r,x))dt, t\geq r,	\\
 	\hspace{1.7mm}Y(r; r,x) = x, 
 	\end{array}
\right.
\end{eqnarray*}
where $(B_t)_{0\leq t\leq T}$ is a standard one-dimensional Brownian motion 
(which is independent of $(c^n_l)_l$), 
$(\mathcal {F}^B_t)_t$ is its Brownian filtration, 
and $b, \sigma : \Bbb {R}\longrightarrow \Bbb {R}$ are Borel functions. 
We assume that $b$ and $\sigma $ are bounded and Lipschitz continuous, that is, 
\begin{eqnarray}\label{Bdd_Lipschitz_Constant}
|\sigma (x) - \sigma (y)| + |b(x) - b(y)| \leq K|x - y|, \ \ 
|\sigma (x)| + |b(x)| \leq K, \ \ x, y\in \Bbb {R} 
\end{eqnarray}
for some $K > 0$. 
Then, for each $r\geq 0$ and $x\in \Bbb {R}$, there exists a unique solution. 

At the end of the time interval $[0,1]$, the trader has an amount of cash $W^n_n$ and 
an amount of the security $\varphi ^n_n$, where 
\begin{eqnarray}\label{W_varphi}
W^n_{l+1} = W^n_l+\psi ^n_l S^n_l e^{-g^n_l(\psi ^n_l)}, \ \ 
\varphi ^n_{l+1} = \varphi ^n_l - \psi ^n_l 
\end{eqnarray}
for $l=0,\ldots ,n-1$ and $W^n_0 = 0,\ \varphi ^n_0 = \Phi _0$. 
We say that an execution strategy $(\psi ^n_l)^{n-1}_{l=0}$ is admissible if 
$(\psi ^n_l)_l\in \mathcal {A}^n_n(\Phi _0)$ holds, where 
$\mathcal {A}^n_k(\varphi )$ is the set of strategies $(\psi ^n_l)^{k-1}_{l=0}$ such that 
$\psi ^n_l$ is $\mathcal {F}_l^n=\sigma \{(B_t)_{t\leq l/n},c^n_0,\cdots , c^n_{l-1}\}$-measurable, 
$\psi ^n_l\geq 0$ for each $l=0,\ldots ,k-1$ 
and $\sum ^{k-1}_{l=0}\psi ^n_l\leq \varphi $ almost surely.

Then, the investor's problem is to choose an admissible strategy to 
maximize the expected utility $\E [u(W^n_n, \varphi ^n_n, S^n_n)]$, where 
$u\in \mathcal {C}$ is the utility function employed by the investor and 
$\mathcal {C}$ is the set of non-decreasing, non-negative, and continuous functions 
on $D = \Bbb {R}\times [0,\Phi _0]\times [0,\infty )$ such that 
\begin{eqnarray}\label{growth_C}
u(w,\varphi ,s)\leq C_u(1+|w|^{m_u}+s^{m_u}), \ \ (w,\varphi ,s)\in D
\end{eqnarray}
for some constants $C_u, m_u>0$. 

For $k=1,\ldots ,n$,\ $(w,\varphi ,s)\in D$ and $u\in \mathcal {C}$, 
we define the discrete-time value function 
$V^n_k(w,\varphi ,s ; u)$ by 
\begin{eqnarray}\nonumber 
V^n_k(w,\varphi ,s ; u) = \sup _{(\psi ^n_l)^{k-1}_{l=0}\in \mathcal {A}^n_k(\varphi )}
\E [u(W^n_k,\varphi ^n_k, S^n_k)]
\end{eqnarray}
subject to (\ref {fluctuate_X}) and (\ref {W_varphi}) for $l=0,\ldots ,k-1$ and 
$(W^n_0,\varphi ^n_0,S^n_0) = (w,\varphi ,s)$ 
(for $s=0$, we set $S^n_l \equiv 0$). 
We denote such a triplet of processes $(W^n_l, \varphi ^n_l, S^n_l)^k_{l=0}$ by 
$\Xi ^n_k(w,\varphi ,s ; (\psi ^n_l)_l)$, 
and denote $V^n_0(w,\varphi ,s ; u) = u(w,\varphi ,s)$. 
Then, this problem is equivalent to consider $V^n_n(0,\Phi _0,s_0 ; u)$. 
We consider the limit of the value function $V^n_k(w,\varphi ,s ; u)$ 
as $n\rightarrow \infty $. 

We introduce the following condition for $g_n(\psi )$, which is also assumed in \cite{Kato}. 
\begin{itemize}
 \item[\mbox {[A]}] \ $\lim _{n\rightarrow \infty }\sup _{\psi \in [0,\Phi _0]}
\left |\frac{d}{d\psi }g_n(\psi )-h(n\psi )\right | = 0$, where 
$h : [0,\infty ) \longrightarrow [0,\infty )$ is a non-decreasing continuous 
function. 
\end{itemize}
Note that in \cite{Kato}, the function $g(\zeta )$ defined by 
\begin{eqnarray}\label{def_g}
g(\zeta ) = \int ^{\zeta}_0h(\zeta ')d\zeta ' 
\end{eqnarray}
represents a MI function in the continuous-time model. 
In our case, $g(\zeta)$ also corresponds to the strength of MI, 
but we still must describe the noise in MI. 

The following are the conditions for $(c^n_l)_l$: 
\begin{itemize}
 \item[\mbox {[B1]}] As a definition, $\gamma _n = \mathop {\rm essinf}_\omega c^n_l(\omega )$.  
For any $n\in \Bbb {N}$, it holds that $\gamma _n > 0$. In addition, 
\begin{eqnarray}\label{cond_B1}
\frac{h(x / \gamma _n)}{n}\longrightarrow 0, \ \ n\rightarrow \infty 
\end{eqnarray}
holds for $x \geq  0$. \vspace{2mm}
 \item[\mbox {[B2]}] 
Let $\mu _n$ be the distribution of $(c^n_0 + \ldots + c^n_{n-1})/n$. 
Then, $\mu _n$ has a weak limit $\mu $ as $n\rightarrow \infty $. 
 \item[\mbox {[B3]}] 
There is a sequence of infinitely divisible distributions $(p_n)_n$ on $\Bbb {R}$ such that 
$\mu _n = \mu *p_n$, and either 
\begin{itemize}
 \item[\mbox {[B3-a]}] $\int _\Bbb {R}x^2p_n(dx) = O(1/n^3)$ as $n\rightarrow \infty $
\end{itemize}
or 
\begin{itemize}
 \item[\mbox {[B3-b]}] There is a sequence $(K_n)_n\subset (0, \infty )$ such that 
$K_n = O(1/n)$, $p_n((-\infty , \allowbreak -K_n))\allowbreak  = 0$ (or $p_n((K_n, \infty )) = 0$) and 
$\int _\Bbb {R}xp_n(dx) = O(1/n)$ as $n\rightarrow \infty $, 
\end{itemize}
where $O$ (Landau's symbol) denotes the order notation. 
\end{itemize}

\begin{remark} \ 
\begin{itemize}
 \item [ (i) ] 
Let us discuss condition [B1]. First, note that $\gamma _n$ is independent of $l$ 
because $c^n_l$, $l = 0, 1, 2, \ldots $ are identically distributed. 
Next, we examine when the convergence (\ref {cond_B1}) holds. 
Since $h$ is non-decreasing, we see that 
\begin{eqnarray*}
\frac{h(x/\gamma _n)}{n} \leq \frac{h(\infty )}{n}, \ \ n\in \Bbb {N}, 
\end{eqnarray*}
where $h(\infty ) =\lim _{\zeta \to \infty}h(\zeta ) \in [0, \infty ]$ 
(which is well-defined by virtue of the monotonicity of $h$). 
This inequality tells us that (\ref {cond_B1}) is fulfilled whenever $h(\infty) < \infty $. 
In the case of $h(\infty ) = \infty $, we have the following example: 
\begin{eqnarray}\label{eg_cond}
h(\zeta ) = \alpha \zeta ^p, \ \ \gamma _n = \frac{1}{n^{1/p - \delta }} \ \ 
(p, \delta > 0, \ \delta \leq  1/p). 
\end{eqnarray}
We can actually confirm (\ref {cond_B1}) by observing that 
\begin{eqnarray*}
\frac{h(x/\gamma _n)}{n} = \frac{\alpha x^p}{n^{p\delta }}\ \longrightarrow \ 0, \ \ 
n\rightarrow \infty . 
\end{eqnarray*}
Note that [B1] always holds when $\inf _n\gamma _n > 0$, 
regardless of whether $h(\infty ) < \infty$. 
 \item [ (ii) ] 
The condition [B2] holds only when $\liminf _n\gamma _n < \infty $. 
Indeed, under [B2] we easily see that the support of the distribution $\mu $ is 
included in the interval $[\liminf _n\gamma _n, \infty )$. 
Note that $\gamma _n$ of (\ref {eg_cond}) satifies $\limsup _n\gamma _n \leq 1$ 
because of the relation $\delta \leq 1/p$. 
 \item [ (iii) ] 
Since $\mu $ is an infinitely divisible distribution, 
there is some L\'evy process (subordinator) $(L_t)_{0\leq t\leq 1}$, defined 
on a certain probability space, such that $L_1$ is distributed according to $\mu $. 
To derive the continuous-time model, 
we want to associate $(c^n_l)_l$ with a difference of $(L_t)_t$, that is, 
to approximate $c^n_l$ from $n(L_{(l+1)/n} - L_{l/n})$. 
The condition [B3] implies that the difference between 
these values is small for large $n$. 
\end{itemize}
\end{remark}

As mentioned in the above remark, there is a L\'evy process $(L_t)_t$ such that 
the distribution of $L_1$ is $\mu $. 
Without loss of generality, we may assume that $(L_t)_t$ and $(B_t)_t$ are defined on the same filtered space. 
Since $(c^n_l)_l$ is independent of $(B_t)_t$, 
we may also assume that $(L_t)_t$ is independent of $(B_t)_t$. 
Let $\nu $ be the L\'evy measure of $(L_t)_t$. 
Since $(L_t)_t$ is a subordinator, 
$\nu $ satisfies $\nu ((-\infty , 0)) = 0$ and either 
\begin{eqnarray}\label{typeA}
\nu ([0, \infty)) < \infty \ \ (\mbox {type A})
\end{eqnarray}
or 
\begin{eqnarray}\label{typeB}
\nu ([0, \infty)) = \infty , \ \ \int _{(0, 1)} z \nu (dz) < \infty \ \ (\mbox {type B}).
\end{eqnarray}
See \cite{Sato} for details. 
Further, we assume the following moment condition for $\nu $: 
\begin{itemize}
 \item[\mbox {[C]}] \ $||\nu ||_1 + ||\nu ||_2 < \infty $, where $||\nu ||_p = \left( \int _{(0, \infty )}z^p\nu (dz)\right)^{1/p} $. 
\end{itemize}
Throughout this paper, we assume [A], [B1]--[B3], and [C]. 

Now, we define the function that gives the limit of the discrete-time value function. 
For $t\in [0,1]$ and $\varphi \in [0,\Phi _0]$ we denote by 
$\mathcal {A}_t(\varphi )$ the set of 
$(\mathcal {F}_r)_{0\leq r\leq t}$-adapted and caglad processes 
(i.e., left-continuous and having a right limit at each point) 
$\zeta  = (\zeta _r)_{0\leq r\leq t}$ 
\vspace{2mm} 
such that 
$\zeta _r\geq 0$ for each $r\in [0,t]$,\ 
$\int ^t_0\zeta _rdr\leq \varphi $ almost surely and 
\begin{eqnarray}\label{def_sup}
||\zeta ||_\infty  := \sup _{(r,\omega )\in [0,t]\times \Omega }\zeta _r(\omega )<\infty , 
\end{eqnarray}
where $\mathcal {F}_r=\sigma \{ B_v, L_v;v\leq r\}\vee \{\mbox{Null sets}\} $. 
Here, the supremum in (\ref {def_sup}) is taken over all values in $[0, t]\times \Omega $. 
Note that we may use the essential supremum in (\ref {def_sup}) in place of the supremum. 

For $t\in [0,1], (w,\varphi ,s)\in D$ and $u\in \mathcal {C}$, 
we define $V_t(w,\varphi ,s ; u)$ by 
\begin{eqnarray}\label{def_conti_v}
V_t(w,\varphi ,s ; u) = \sup _{(\zeta _r)_{r}\in \mathcal {A}_t(\varphi )}
\E [u(W_t,\varphi _t, S_t)] 
\end{eqnarray}
subject to 
\begin{align}\nonumber 
dW_r &= \zeta _rS_rdr, \\\nonumber 
d\varphi _r &= -\zeta _rdr, \\
dX_r &= \sigma (X_r)dB_r+b(X_r)dr-g(\zeta _r)dL_r, 
\label{SDE_X_g}\\\nonumber 
S_r &= \exp (X_r) 
\end{align}
and $(W_0,\varphi _0,S_0) = (w,\varphi ,s)$. 
We denote such a triplet of processes 
$(W_r, \varphi _r, S_r)_{0\leq r\leq t}$ by $\Xi _t(w,\varphi ,s ; (\zeta _r)_r)$. 
Note that $V_0(w,\varphi ,s ; u) = u(w,\varphi ,s)$. 
We call $V_t(w,\varphi ,s ; u)$ a continuous-time value function. 
Also note that $V_t(w,\varphi ,s ; u)<\infty $ for each $t\in [0,1]$ and 
$(w,\varphi ,s)\in D$.

\begin{remark}
Condition [C] guarantees that the 
SDE (\ref {SDE_X_g}) has a unique solution for each given $(\zeta _r)_r\in \mathcal {A}_t(\varphi )$ 
(from Theorem 1.19 in \cite {Oksendal-Sulem}; note that 
the finiteness of $||\nu ||_1$ is required for uniqueness). 
Moreover, by Lemma \ref{lem_comparison} in Section \ref {sec_proof}, we can show that 
\begin{eqnarray*}
0\leq S_r \leq \exp \left( Y(r ; 0, \log s)\right) , \ \ r\in [0, t] \ \ \mbox {a.s.}, 
\end{eqnarray*}
so that, 
applying Lemma \ref {Lemma_Moment_Estimate}, 
for each $m > 0$, 
\begin{eqnarray}\label{moment}
\E \left [\sup _{r\in [0, t]}|W_r|^{m}\right ] + 
\E \left [\sup _{r\in [0, t]}|S_r|^{m}\right ] \leq C_{m,K,\Phi_0}(|w|^{m} + s^{m})
\end{eqnarray}
for some $C_{m,K,\Phi_0} > 0$, where $K>0$ is as given in (\ref{Bdd_Lipschitz_Constant}). 
\end{remark}

Now we give the convergence theorem for value functions. 

\begin{theorem} \ \label{converge_random}
For each $(w,\varphi ,s)\in D$, $t\in [0,1]$ and $u\in \mathcal {C}$ it holds that 
\begin{eqnarray*}
\lim _{n\rightarrow \infty }V^n_{[nt]}(w,\varphi ,s;u) = V_t(w,\varphi ,s;u), 
\end{eqnarray*}
where $[nt]$ is the greatest integer smaller than or equal to $nt$. 
\end{theorem}

According to this theorem, a discrete-time value function converges to \\
$V_t(w, \varphi ,s \allowbreak ; u)$ 
by shortening the time intervals of execution. 
This implies that we can regard $V_t(w, \varphi ,s ; u)$ as the value function of 
the continuous-time model of an optimal execution problem with random MI. 
This result is almost the same as in \cite {Kato}, 
with the exception that the term of MI is given as an increment $g(\zeta _r)dL_r$. 
Let 
\begin{eqnarray*}
L_t = \gamma t + \int ^t_0\int _{(0, \infty )}zN(dr, dz)
\end{eqnarray*}
be the L\'evy decomposition of $(L_t)_t$, 
where $\gamma \geq 0$ and $N(\cdot, \cdot)$ is a Poisson random measure 
(see \cite {Papapantoleon, Sato}, for instance). 
Then, $g(\zeta _r)dL_r$ can be divided into two terms as follows: 
\begin{eqnarray*}
g(\zeta _r)dL_r = \gamma g(\zeta _r)dr + g(\zeta _r)\int _{(0, \infty )}z N(dr, dz). 
\end{eqnarray*}
The last term on the right side indicates the effect of noise in MI. 
This means that noise in MI appears as a jump of a L\'evy process. 
Using the above representation and It\^o's formula, 
we see that when $s > 0$ the process $(S_r)_r$ satisfies 
\begin{eqnarray*}
dS_r = \hat{\sigma }(S_r)dB_r+\hat{b}(S_r)dr - 
\left\{ \gamma g(\zeta _r)S_r dr + S_{r-}\int _{(0, \infty )}(1 - e^{-g(\zeta _r)z})N(dz, dr)\right\} ,
\end{eqnarray*}
where $\hat{\sigma }(s) = s\sigma (\log s)$ and 
$\hat{b}(s) = s\Big\{ b(\log s)+\frac{1}{2}\sigma (\log s)^2\Big\} $ for $s > 0$ 
(with \\$\hat{\sigma }(0) = \hat{b}(0) = 0$). 

\begin{remark} 
It is well known that MI can be divided into 
two parts: a permanent part and a temporary (or transient) part 
(see \cite{Almgren-Chriss,Holthausen-et-al} and others). 
In our study, we mainly treat the permanent MI and 
do not model the temporary one for the same reason as in Remark 2 in \cite {Kato}. 
However, as in \cite {Kato2}, we can introduce a price recovery effect 
by considering, for instance, an Ornstein--Uhlenbeck (OU)-type process such as 
\begin{eqnarray}\label{OU_type}
dX_r = \beta (F_r - X_r)dr + \sigma (X_r)dB_r - g(\zeta _r)dL_r, 
\end{eqnarray}
where $\beta > 0$ denotes the speed of price recovery and 
$(F_r)_r$ is a log-fundamental value process 
(\cite{Kato2} studies the case where $F_r = \mathrm {Const.}$ and $dL_r = dr$). 
Then, we can implicitly consider the transient MI in our model. 
Properties of optimal strategies under the log-price process (\ref {OU_type}) 
are studied in \cite {Ishitani-Kato_JSIAM} for the case 
where we restrict the admissible strategies to deterministic ones and 
$F_r = \mathrm {Const}$. 
We leave the case of adaptive strategies as an area for future study. 
\end{remark}

\begin{remark}
$g$ describes the shape of the MI function, and 
assumption [A] implies that $g$ is convex in the wide sense. 
In practice it is said that the natural form of MI functions is ``S-shaped,'' 
that is, concave for small selling and convex for large selling \cite {Kato_JSIAM}. 
In this case, the derivative $h = g'$ of the MI function is no longer monotonous. 
Derivation of an optimal execution problem with an S-shaped deterministic MI function 
is studied in \cite {Kato_JSIAM}. In the case of random MI, 
we further require that $\gamma $ is strictly positive for technical reason. 
For details see \cite {Ishitani-Kato_JSIAM}, 
in which we study the discrete approximation of the continuous-time value function 
with random MI functions. 
\end{remark}

\begin{remark}
Theorem \ref{converge_random} 
has the same assertion as Theorem A.1 in \cite {Kato}, 
and the outlines of our proofs are based on those of \cite {Kato}. 
However demonstrating our theorems requires a significant improvement of the proofs. 
In particular, it is hard to show the $L^2$ convergence of controlled processes because of 
a technical difficulty caused by the jump term of $(L_t)_t$. 
To overcome this problem, we prepare a useful lemma (Lemma \ref {Ishi_Lem} in Section \ref {prelim}) and 
we give the proofs by properly using both $L^1$ and $L^2$ moments to see the convergences of the processes. 
This is one of the mathematical contributions of this paper. 
See Section \ref {subsec_conv} and 
for details. 
See also Remark 2.2(i) in \cite{Ishitani-Kato_COSA2}. 
\end{remark}

\section{Proofs} \label{sec_proof}

In this section we prepare several lemmas that we use to prove 
Theorems \ref {converge_random}. 
Our approach for the proof is similar to those adopted by \cite {Kato}. 

\subsection{Preliminaries} \label{prelim}

\begin{lemma} \label{lemm_conti_u} 
Let $\Gamma _k$ $(k\in \Bbb {N})$ be sets, $u\in {\mathcal {C}}$, and let 
$(W^i(k,\gamma ), \varphi ^i(k,\gamma ), S^i(k,\gamma ))\in D$ 
$(\gamma \in \Gamma _k$, $k\in \Bbb {N}$, $i=1,2)$ be random variables. 
Assume that 
\begin{eqnarray*}
&&\lim _{k\rightarrow \infty }\sup _{\gamma \in \Gamma _k} 
\E [|W^1(k,\gamma )-W^2(k,\gamma )|^{m_1} + |\varphi ^1(k,\gamma )-\varphi ^2(k,\gamma )|^{m_2}\\
&&\hspace{57mm} + |S^1(k,\gamma )-S^2(k,\gamma )|^{m_3}] = 0
\end{eqnarray*}
and 
\begin{eqnarray*}
\sum ^2_{i=1}\sup _{k\in \Bbb {N}}\sup _{\gamma \in \Gamma _k}
\E [|W^i(k,\gamma )|^{m_4}+(S^i(k,\gamma ))^{m_4}] < \infty 
\end{eqnarray*}
for some $m_1, m_2, m_3 > 0$ and $m_4 > m_u$, where $m_u$ is as appeared in $(\ref {growth_C})$.
Then we have
\begin{eqnarray*}
&&\lim _{k\rightarrow \infty }\sup _{\gamma \in \Gamma _k}
\big| \E [u(W^1(k,\gamma ), \varphi ^1(k,\gamma ), S^1(k,\gamma ))] \nonumber \\
&&\qquad \qquad \quad -  \E [u(W^2(k,\gamma ), \varphi ^2(k,\gamma ), S^2(k,\gamma ))]\big| = 0 .
\end{eqnarray*}
\end{lemma}

The above lemma is a generalization of Lemma B.2 in \cite {Kato}.
One can prove Lemma \ref {lemm_conti_u} by using 
the H\"older inequality, the Chebyshev inequality, 
and uniform continuity of $u(w,\varphi ,s)$ on any compact set.

Here, we quote Lemma B.1 in \cite {Kato}, as follows, because 
we frequently use this lemma in the proofs: 
\begin{lemma}\label{Lemma_Moment_Estimate} 
Let $Z(t; r, s) = \exp (Y(t; r, \log s))$ and 
$\hat{Z}(s) = \sup_{0\leq r\leq 1}Z(r; 0, s)$. 
Then, for each $m>0$, there is a constant 
$C_{m, K}>0$ depending only on $K$ and $m$ such that 
$E [\hat{Z}(s)^m]\leq C_{m, K} s^m$, 
where $K>0$ is a constant appearing in (\ref{Bdd_Lipschitz_Constant}).
\end{lemma}

\noindent
\begin{lemma} \label{Ishi_Lem}
Let $(X^{k, i}_r)_{r\in [0,1]}$, $i = 1, 2, k \in \Bbb {N}$, 
be $\mathbb{R}$-valued $(\mathcal {F}_r)_r$-progressive processes satisfying 
\begin{eqnarray*}
X^{k, i}_r = x^{k, i} + \int ^r_0b(X^{k, i}_v)dv + \int ^r_0\sigma (X^{k, i}_v)dB_v + F^{k, i}_r , 
\ \ r\in [0,1],
\end{eqnarray*}
with $x^{k, i} \in \Bbb{R}$  
for $i = 1, 2$ and $k \in \Bbb {N}$, where 
$(F^{k, i}_r)_r$ are $(\mathcal {F}_r)_r$-adapted processes of bounded variation, 
and let $\Pi_k \subset [0,1]$, $k \in \Bbb {N}$, be Borel sets. 
Moreover, assume that 
\begin{description}
 \item[(i)] $x^{k, 1} - x^{k, 2}\longrightarrow 0, \ \ k\rightarrow \infty $, 
 \item[(ii)] 
$\lim_{k\to \infty} \left\{ D^k_1+\int_0^1 D^k_r dr \right\}= 0$,  
where 
\begin{eqnarray*}
D^k_r = \E \left [\sup_{v\in \Pi_k(r) }|F^{k, 1}_v - F^{k, 2}_v|\right ], \ \ 
\Pi_k(r)=([0,r]\cap \Pi_k)\cup \{r\}. 
\end{eqnarray*}
\end{description}
Then it holds that 
\begin{eqnarray*}
\E \left [\sup_{v \in \Pi_k } \left| X^{k, 1}_v - X^{k, 2}_v \right| \right ]  
\longrightarrow 0, \ \ k \rightarrow \infty . 
\end{eqnarray*}
\end{lemma}

\begin{proof} \ 
Define $(\tilde{X}^k_r)_r$ by 
\begin{eqnarray*}
\tilde {X}^k_r = x^{k, 2} + 
\int ^r_0 b(X^{k, 1}_v) dv + \int ^r_0\sigma (X^{k, 1}_v)dB_v + F^{k, 2}_r 
\end{eqnarray*}
and let 
\begin{eqnarray*}
\tilde{D}^k_r = 
\E \left[ \sup _{v\in \Pi_k(r)}|\tilde {X}^k_v -  X^{k, 1}_v| \right], \ \ 
\Delta ^k_r = 
\E\left[ \sup _{v\in \Pi_k(r)} \left\vert 
\tilde {X}^k_v - X^{k, 2}_v
\right\vert ^2 \right]. 
\end{eqnarray*}
Note that $\Delta ^k_r$ is finite because of the boundedness of $b$ and $\sigma $. 
We deduce that 
\begin{eqnarray}\label{temp_001}
\E [\sup_{v \in \Pi_k(r) } \left| X^{k, 1}_v - X^{k, 2}_v \right| ] 
\leq \tilde{D}^k_r + (\Delta ^k_r)^{1/2},\ r\in [0,1]. 
\end{eqnarray}
Combining the obvious inequality $\tilde{D}^k_r \leq \vert x^{k, 1}-x^{k, 2}\vert + D^k_r$ 
with (i) and (ii), we see that 
\begin{eqnarray}\label{temp_003}
\tilde{D}^k_1 + \int _0^1 \tilde{D}^k_r dr \ \longrightarrow \ 0, \ \ 
k\rightarrow \infty . 
\end{eqnarray}
Moreover, applying Doob's maximal inequality and the Schwarz inequality, we have that 
\begin{eqnarray}
\Delta ^k_r\leq 
8\E \left[ \int_0^r \left\{ 
\left| \sigma (X^{k, 1}_v) - \sigma (X^{k, 2}_v) \right| ^2 + 
\left| b(X^{k, 1}_v) - b(X^{k, 2}_v) \right| ^2 \right\} dv \right ] . 
\label{Ishi_Lem_ieq1}
\end{eqnarray} 
Then we observe that
\begin{eqnarray*}
&&|\sigma (X^{k, 1}_v) - \sigma (X^{k, 2}_v)|^2 \\
&&\quad \leq 4 K^2 \{ 1_{\Omega _k(v)} + |\tilde{X}^k_v - X^{k, 1}_v|1_{\Omega _k(v)^c} + 
|\tilde{X}^k_v - X^{k, 2}_v|^21_{\Omega _k(v)^c} \} 
\end{eqnarray*}
to arrive at 
\begin{align}
& \E \left[ \int_0^r 
\left| \sigma (X^{k, 1}_v) - \sigma (X^{k, 2}_v) \right| ^2 dv \right]  
\nonumber \\
&\quad \leq  
4K^2\left\{ \int_0^r P(\Omega _k(v))dv + \int_0^r \tilde{D}^k_v dv + \int ^r_0\Delta ^k_vdv \right\} 
\nonumber \\
&\quad \leq 8 K^2 \left\{ \int_0^r \tilde{D}^k_v dv + \int ^r_0\Delta ^k_vdv \right\}  
\label{est_sigma}
\end{align}
by using the Chebyshev inequality, 
where 
\begin{eqnarray*}
\Omega _k(r) := \{\sup _{v\in \Pi_k(r)}|\tilde {X}^k_v -  X^{k, 1}_v| > 1 \} . 
\end{eqnarray*}
Similarly, we get 
\begin{eqnarray}\label{est_b}
\E \left[ \int_0^r 
\left| b(X^{k, 1}_v) - b(X^{k, 2}_v) \right| ^2 dv \right]  \leq 
8K^2\left\{ 
\int_0^r \tilde{D}^k_v dv
+ \int ^r_0\Delta ^k_vdv \right\} . 
\end{eqnarray}
Combining (\ref {est_sigma}) and (\ref {est_b}) with (\ref {Ishi_Lem_ieq1}), we get 
\begin{eqnarray*}
\Delta ^k _r \leq  128 K^2 \left\{ 
\int_0^1 \tilde{D}^k_v dv
+ \int^r_0 \Delta ^k _vdv \right\} . 
\end{eqnarray*}
Applying the Gronwall inequality, we deduce that 
\begin{eqnarray}\label{temp_002}
\Delta ^k_r \leq C\int_0^1 \tilde{D}^k_v dv, \quad r \in [0,1]
\end{eqnarray}
for some $C > 0$. 
Our assertion is now obtained from (\ref {temp_001}), (\ref {temp_003}), and (\ref {temp_002}). 
\end{proof}

We can obtain the following lemma, which we need to prove Theorem \ref{converge_random} 
by a standard argument. 
\begin{lemma}\label{Lemma_Moment_Uniform_Continuity} 
Let $t\in [0,1]$, $\varphi \geq 0$, $x\in \Bbb{R}$ and 
$(\zeta_r)_{0\leq r\leq t}\in \mathcal{A}_t(\varphi)$. 
Assume further that $(X_r)_{0\leq r\leq t}$ is given by $(\ref {SDE_X_g})$ 
with $(\zeta _r)_r$ and $X_0 = x$. Then, we have 
\begin{eqnarray*}
E \left[\sup_{r\in [r_0,r_1]}
\left|X_r-X_{r_0}+\int^r_{r_0}g(\zeta_v)dL_v\right|^{2p}\right] 
\leq \widetilde{C}_{p, K} (r_1-r_0)^p
\end{eqnarray*}
for $p>0$ and $0\leq r_0\leq r_1\leq t$, 
where $K>0$ is a constant appearing in (\ref{Bdd_Lipschitz_Constant}) 
and $\widetilde{C}_{p, K}>0$ depend only on $p$ and $K$. 
\end{lemma}

Arguments similar
to the proof of Proposition 5.2.18 in \cite{KarazasShreve} lead us to the following lemma:
\begin{lemma}\label{lem_comparison} \ 
Let $t\in [0,1]$, $\varphi \geq 0$, $x\in \Bbb {R}$, 
$(\zeta _r)_{0\leq r\leq t}, (\zeta '_r)_{0\leq r\leq t}
\in \mathcal {A}_t(\varphi )$ and suppose 
$(X_r)_{0\leq r\leq t}$ $($resp., $(X'_r)_{0\leq r\leq t}$$)$ 
is given by $(\ref {SDE_X_g})$ with $(\zeta _r)_r$ 
$($resp., $(\zeta '_r)_r$$)$ and $X_0 = x \leq X'_0$. 
Suppose $\zeta _r\leq \zeta '_r$ for any $r\in [0,t]$ almost surely. 
Then $X_r\geq X'_r$ for any $r\in [0,t]$ almost surely. 
\end{lemma}

Note that the above lemma itself can be proved without finiteness of $||\nu ||_2$. 

\subsection{Proof of Theorem \ref {converge_random} }\label{subsec_conv}

From [B2] and [B3], we see that there exists a L\'evy process $(Z^n_t)_t$ 
which is independent of $(L_t)_t$ and $(B_t)_t$, and that the distribution of $Z^n_1$ is $p_n$. 
Then, the stochastic process $L^n_t = L_t + Z^n_t$ also becomes a L\'evy process. 
Now, define $(\tilde{c}^n_k)_k$ by 
\begin{eqnarray*}
\tilde{c}^n_l = n(L^n_{(l+1)/n} - L^n_{l/n}). 
\end{eqnarray*}
Then $(\tilde{c}^n_k)_k$ are i.i.d.\hspace{1mm}random variables with the same distribution as $(c^n_l)_l$. 
Therefore, $V^n_k(w, \varphi , s ; u)$ coincides with $\tilde{V}^n_k(w, \varphi , s ; u)$, 
where $\tilde{V}^n_k(w, \varphi , s ; u)$ is the value function defined 
as the same way as $V^n_k(w, \varphi , s ; u)$, replacing $(c^n_l)_l$ with $(\tilde{c}^n_l)_l$. 
Thus we can identify $(c^n_l)_l$ and $(\tilde{c}^n_l)_l$ without loss of generality 
(similarly, $\mathcal {F}^n_l$ is identified as $\mathcal {F}_{l/n}$). 
Let 
\begin{align}
C^* &:= \sup_n\left( 
n \max_{l=0,\cdots ,n-1}\E [ | n L_{(l+1)/n}-n L_{l/n}-\tilde{c}^n_l | ] \right) 
\nonumber \\
&=
\sup_n n^2\E [ |Z^n_{1/n}| ] < \infty . \label{Isshi_condition_c_l}
\end{align}
Here, the finiteness of $C^*$ comes from [B3] and the following relations: 
\begin{align*}
\E [(Z^n_{1/n})^2] &= \frac{1}{n} 
\int _\Bbb {R}x^2 p_n(dx) -\frac{n-1}{n^2}\left(\int _\Bbb {R}x p_n(dx)\right)^2, \\
\E [|Z^n_{1/n}|] &= \frac{1}{n} 
\int _\Bbb {R}x p_n(dx)+2 \E [(-Z^n_{1/n})1_{[0, \infty)}(-Z^n_{1/n})]\\
&=
-\frac{1}{n} \int _\Bbb {R}x p_n(dx)+2 \E [Z^n_{1/n}1_{[0, \infty)}(Z^n_{1/n})]. 
\end{align*}
Note that the function $g_n$ on $[0, \Phi _0]$ can be extended on 
$[0, \infty)$ by 
\begin{eqnarray*}
g_n(\psi)=g_n(\Phi _0)+\int_{\Phi _0}^{\psi} h(n \psi ')d\psi ', \ \ 
\psi \in [\Phi_0, \infty ). 
\end{eqnarray*}

We can now give a proof of Theorem \ref {converge_random}. 
We divide the proof into the following two propositions. 

\begin{proposition}\label{thm1_pro1} 
$\limsup _{n\rightarrow \infty }V^n_{[nt]}(w,\varphi ,s;u) \leq V_t(w,\varphi ,s;u)$.
\end{proposition}
\begin{proposition}\label{thm1_pro2} 
$\liminf _{n\rightarrow \infty }V^n_{[nt]}(w,\varphi ,s;u) \geq V_t(w,\varphi ,s;u)$.
\end{proposition}

\begin{proof}[Proof of Proposition \ref {thm1_pro1}]
For brevity, we assume that $t=1$. 
First of all, analogously to the proof of Proposition B.24 in \cite {Kato}, 
we can show that there exists an optimal strategy 
$(\hat{\psi }^n_l)^{n-1}_{l=0}\in \mathcal {A}^n_n(\varphi )$ corresponding to 
the value function $V^n_n(w, \varphi , s; u)$ such that 
$0\leq \hat{\psi }^n_l\leq \min \{ \psi ^*_n, \Phi _0\} $ for each $l = 0, \ldots , n - 1$, 
where 
\begin{eqnarray*}
\psi ^*_n=\sup \{\psi \geq 0 \ ; \ \gamma _n\psi h_n(\psi ) \leq 1\}, 
\end{eqnarray*}
$h_n = g_n'$, and $\gamma _n$ is given in [B1]. 
Set 
\begin{eqnarray*}
c^*_n=\left\{ \begin{array}{ll}
2 \varepsilon _n + h_n(\psi ^*_n) & (h(\infty)=\infty), \\
\varepsilon _n + h(\infty) & (h(\infty)<\infty), 
\end{array} \right.
\end{eqnarray*}
where 
\begin{eqnarray*}
\varepsilon _n 
= \sup _{\psi \geq 0}
\left |\frac{d}{d\psi }g_n(\psi )-h(n\psi )\right |. 
\end{eqnarray*} 
Then we can prove that 
\begin{eqnarray}\label{conv_c*}
\frac{c^*_n}{n} \ \longrightarrow \ 0, \ \ n\rightarrow \infty . 
\end{eqnarray}
Indeed, when $h(\infty)<\infty$, (\ref{conv_c*}) is obvious from [A]. 
When $h(\infty) = \infty$, if (\ref {conv_c*}) is not true, 
we see that for each $M > 0$ there is an increasing sequence $(n_k)_k\subset \Bbb {N}$ 
such that $n_k\gamma _{n_k}\psi ^*_{n_k}\leq M$, $k\in \Bbb {N}$ 
(for brevity we omit $k$ in the notations below). 
Then we have that $\psi ^*_n\leq M / (n\gamma _n)$ and that 
\begin{eqnarray*}
h_n(\psi ^*_n) \leq \varepsilon _n + h(n\psi ^*_n) 
\leq \varepsilon _n + h\left( \frac{M}{\gamma _n}\right) .
\end{eqnarray*}
Since $h(\infty)=\infty$ and $\lim_{n\to \infty} \varepsilon _n = 0$, 
it holds that $\gamma _n \psi ^*_n h_n(\psi ^*_n ) = 1$ for a sufficiently large $n$, thus 
\begin{eqnarray*}
1 \leq \frac{M}{n}\left\{  \varepsilon _n + h\left( \frac{M}{\gamma _n}\right) \right\} . 
\end{eqnarray*}
However, [B1] implies that the right side of the above inequality converges to zero 
as $n\rightarrow \infty $, which leads to a contradiction. 
Therefore, we get (\ref {conv_c*}) and see that 
\begin{eqnarray}\label{est_opt}
g_n(\hat{\psi }^n_l) = \int _0^{\hat{\psi }^n_l} h_n(\psi ') d\psi '  \leq c^*_n\hat{\psi }^n_l, \ \ l=0,\ldots ,n-1. 
\end{eqnarray}
\begin{remark} \ 
\begin{itemize}
 \item [ (i) ] In \cite {Kato}, the left side of (\ref {est_opt}) is bounded from above uniformly in $n$. 
However, we cannot show the same inequality in our case because of the noise of MI function $g^n_l$. 
 \item [ (ii) ] 
When $\inf _n\gamma _n > 0$, we can show the uniform boundedness of the left side of (\ref {est_opt}). 
\end{itemize}
\end{remark}

To continue the proof of Proposition \ref {thm1_pro1}, 
we construct the continuous-time strategy 
$(\hat{\zeta }_r)_r\in \mathcal {A}_1(\varphi )$ by 
$\hat{\zeta } _0 = 0$ and 
$\hat{\zeta } _r=n\hat{\psi} ^n_{\lceil nr - 1\rceil }\ (r>0)$, 
where \\$\lceil x \rceil :=\min \{ n\in \Bbb{Z} ; x\leq n\}$ 
is the ceiling function. 
Let $(W^n_l, \varphi ^n_l, S^n_l)^{n-1}_{l=0} = \Xi ^n_n(w, \varphi , s ; (\hat{\psi }^n_l)_l)$ and 
$(W_r, \varphi _r, S_r)_{0\leq r\leq 1} = \Xi _1(w, \varphi  , s ; (\hat{\zeta }_r)_r)$, 
and let $X^n_l = \log S^n_l$ and $X_r = \log S_r$. 

Our first step is to apply Lemma \ref {Ishi_Lem} with 
\begin{eqnarray*}
F^{n, 1}_r = - \int_0^{r} g(\hat{\zeta }_{v}) dL_{v}, \ \ 
F^{n, 2}_r = - \sum_{l=0}^{\lceil nr - 1\rceil }g^n_l(\hat{\psi }^n_l)
\end{eqnarray*}
and $\Pi_n = \{ l/n ; l = 0,\cdots , n \}$ to obtain 
\begin{eqnarray}\label{IshiLem_result_1}
\E \left [\max _{l = 0, \ldots , n} | X_{l/n} - X^n_l | \right ] =
\E \left [\sup _{v\in \Pi_n}|X_v - \tilde{X}^n_v |\right ]\longrightarrow 0 , \ \ 
n\rightarrow \infty , 
\end{eqnarray}
where we denote $\sum ^{-1}_{l=0} = 0$ 
and $(\tilde{X}^n_r)_{r\in [0,1]}$ is given by 
\begin{eqnarray}\label{def_of_tildeX}
\tilde{X}^n_r = Y\Big (r ; \frac{k}{n}, X^n_k-g^n_k(\hat{\psi }^n_k)\Big ),\ \ 
r\in \left (\frac{k}{n}, \frac{k+1}{n}\right ] 
\end{eqnarray}
and $\tilde{X}^n_0 = \log s$. 
Note that $(\tilde{X}^n_r)_r$ satisfies $\tilde{X}^n_{l/n} = X^n_l$ for 
$l = 0, \ldots , n$ and 
\begin{eqnarray*}
\tilde{X}^n_r = \log s + \int ^r_0\sigma (\tilde{X}^n_v)dB_v + 
\int ^r_0b(\tilde{X}^n_v)dv + F^{n, 2}_r. 
\end{eqnarray*}
To apply Lemma \ref {Ishi_Lem}, 
it suffices to show that 
\begin{align}
\E \left [\sup _{v\in \Pi_n} | F^{n, 1}_v - F^{n, 2}_v |\right ] + 
\int _0^1 \E \left [\sup _{v\in \Pi_n(r)} | F^{n, 1}_v - F^{n, 2}_v |\right ] dr \longrightarrow 0 , \ \ 
n\rightarrow \infty . 
\label{Prop1_1_condition_for_IshiLem_0}
\end{align}
A straightforward calculation gives 
\begin{align}
\sup_{v\in \Pi_n(r)}| F^{n, 1}_v - F^{n, 2}_v |
\leq & 
\sum_{l=0}^{n-1}\left |\frac{1}{n}g(n \hat{\psi }^n_l) - g_n(\hat{\psi }^n_l) \right| n (L_{(l+1)/n}-L_{l/n}) 
\nonumber \\
& + \sum_{l=0}^{n-1} g_n(\hat{\psi }^n_l) \left | nL_{(l+1)/n}-nL_{l/n}-\tilde{c}^n_l \right | 
\label{Prop1_1_condition_for_IshiLem_1} \\
& + 1_{[0, 1]\setminus \Pi _n}(r) g(n \hat{\psi }^n_{\lceil nr \rceil }) (L_r - L_{\lceil nr -1 \rceil/n}) , 
\quad r \in [0, 1].
\nonumber 
\end{align}
From the independence of $\hat{\psi }^n_l$ and $L_{(l+1)/n}-L_{l/n}$ and 
\begin{eqnarray*}
\sup_{\psi \in (0, \Phi_0]}
\left |\frac{g(n \psi )}{n\psi } - \frac{g_n(\psi )}{\psi } \right| \leq \varepsilon _n 
\longrightarrow 0, \ \ \varepsilon \rightarrow 0, 
\end{eqnarray*}
we have that 
\begin{eqnarray}
\E\left [
\sum_{l=0}^{n-1}\left |\frac{1}{n}g(n \hat{\psi }^n_l) - g_n(\hat{\psi }^n_l) \right| n (L_{(l+1)/n}-L_{l/n})
\right ]
\leq \tilde{\gamma }\Phi_0 \varepsilon _n  
\longrightarrow 0 , \ \ 
n\rightarrow \infty . \label{Prop1_1_condition_for_IshiLem_2} 
\end{eqnarray}
Also, from (\ref{Isshi_condition_c_l}) and 
the independence of $\hat{\psi }^n_l$ and $(L_{(l+1)/n}-L_{l/n}, \tilde{c}^n_l)$, we see that 
\begin{eqnarray}\label{Prop1_1_condition_for_IshiLem_3} 
\E\left [\sum^{n-1}_{l=0}g_n(\hat{\psi }^n_l) 
\left | nL_{(l+1)/n}-nL_{l/n}-\tilde{c}^n_l \right | \right ]
\leq \frac{c^*_n}{n}C^*\Phi_0 
\longrightarrow 0 , \ \ n\rightarrow \infty .
\end{eqnarray}
On the other hand, from the independence of $\hat{\psi }^n_{\lceil nr \rceil }$ 
and $L_r - L_{\lceil nr -1 \rceil/n}$, we can obtain that
\begin{align*}
&\E\left [ 1_{[0, 1]\setminus \Pi _n}(r) 
g(n \hat{\psi }^n_{\lceil nr \rceil })  (L_r - L_{\lceil nr -1 \rceil/n}) \right ] \\
&\quad \leq \E\left [
( c^*_n \hat{\psi }^n_{\lceil nr \rceil } + \Phi_0 \varepsilon_n )
n (L_r - L_{\lceil n r -1 \rceil/n}) \right ] \\
&\quad = \tilde{\gamma }( c^*_n \E[ \hat{\psi }^n_{\lceil nr \rceil } ] + \Phi_0 \varepsilon_n ) 
(n r - \lceil n r -1 \rceil), 
\end{align*}
hence 
\begin{align}\label{Prop1_1_condition_for_IshiLem_4} 
&\int_0^1 \E\left [ 1_{[0, 1]\setminus \Pi _n}(r) 
g(n \hat{\psi }^n_{\lceil nr \rceil })  (L_r - L_{\lceil nr -1 \rceil/n}) \right ] dr 
\nonumber \\
&\quad \leq \tilde{\gamma }\Phi_0 
\left\{ \frac{c^*_n}{n} + \varepsilon_n \right\}
\longrightarrow 0 , \ \ n\rightarrow \infty .
\end{align}
By combining 
(\ref{Prop1_1_condition_for_IshiLem_1})--(\ref{Prop1_1_condition_for_IshiLem_4}) 
we can prove 
(\ref{Prop1_1_condition_for_IshiLem_0}), 
and thus we obtain $(\ref{IshiLem_result_1})$. 

Using the monotonicity of $u$, we observe that
\begin{align}\label{temp_step_1_0}
V^n_n(w,\varphi ,s ; u)-V_1(w,\varphi ,s ; u) &\leq 
\E [u(W^n_n,\varphi ^n_n,S^n_n)] - \E [u(W_1, \varphi _1, S_1 )] \nonumber \\
&\leq 
\E[u(W^n_n,\varphi^n_n,S^n_n)] - \E [u(\ddot {W}^n_n,\varphi^n_n,S^n_n)] \\
& \quad + 
\E [u(\tilde{W}^n_n,\varphi ^n_n,S^n_n)] - \E [u(W_1, \varphi _1, S_1)], \nonumber 
\end{align}
where 
\begin{align*}
\ddot{W}^n_n &= w + \sum _{l=0}^{n-1}  \hat{\psi }^n_l 
\exp (X^n_l-n(L_{(l+1)/n}-L_{l/n})g_n(\hat{\psi }^n_l)) , \\
\tilde{W}^n_n &= 
w + \sum _{l=0}^{n-1} n \hat{\psi }^n_l \int_{l/n}^{(l+1)/n} 
\exp (X^n_l-n(L_{r}-L_{l/n})g_n(\hat{\psi }^n_l))dr. 
\end{align*}
Note that $\tilde{W}^n_n \geq \ddot{W}^n_n $ holds almost surely.

From (\ref{Prop1_1_condition_for_IshiLem_3}), Lemma \ref{Lemma_Moment_Estimate}, 
and the inequality 
\begin{align}\label{Inequality_Diff_ex_and_ey}
\vert e^x -e^y \vert \leq (e^x + e^y) \vert x - y\vert, \quad x, y\in \mathbb{R},
\end{align}
we can obtain that
\begin{align}\nonumber 
\E[ |\ddot{W}^n_n - W^n_n|^{1/2} ] 
&\leq \widehat{C} \sqrt{\Phi_0} 
\E\Big[ \sum_{l=0}^{n-1} g_n(\hat{\psi }^n_l ) 
\left\vert n(L_{(l+1)/n}-L_{l/n}) - \tilde{c}^n_l \right\vert \Big] \\
&\longrightarrow 0 , \ \ n\rightarrow \infty , 
\label{conv_ddotW_W}
\end{align}
where $\widehat{C} = (2sC_{1, K})^{1/2}$ and $C_{1, K}$ is given in Lemma \ref {Lemma_Moment_Estimate}. 
Further, applying Lemma \ref{Lemma_Moment_Uniform_Continuity} and using (\ref{IshiLem_result_1}) 
and (\ref{Inequality_Diff_ex_and_ey}), we see that 
\begin{align}\nonumber 
&\E[ |\tilde{W}^n_n - W_1|^{1/2} ]\\\nonumber 
&\quad \leq \widehat{C} \sqrt{\Phi_0} \E \Big[ 
\sup_{l=0,\ldots , n-1} \sup_{r \in [l/n, (l+1)/n]}
\big\vert X_r-X^n_l+n(L_r-L_{l/n}) g_n(\hat{\psi }^n_l ) \big\vert
\Big] ^{1/2}\nonumber \\
&\quad \leq \widehat{C} \sqrt{\Phi_0} \Big\{ \widetilde{C}_{2, K} \frac{1}{n^{1/4}}
 + \E \Big[\max _{l = 0, \ldots , n} | X_{l/n} - X^n_l | \Big]  \Big\}^{1/2}
\longrightarrow 0 , \ \ n\rightarrow \infty . \label{conv_tildeW_W}
\end{align}
Moreover, obviously it holds that $\varphi ^n_n = \varphi _1$ and 
\begin{align}\label{convergence_Snn_S1}
\E [\vert S_1 - S^n_n \vert^{1/2} ]\leq 
\widehat {C}\E[ \vert X_1-X^n_n \vert ] ^{1/2}
\longrightarrow 0 , \ \ n\rightarrow \infty .
\end{align}
From (\ref{conv_ddotW_W})--(\ref{convergence_Snn_S1}), 
we can apply Lemma \ref {lemm_conti_u} to see that 
\begin{eqnarray}
\label{temp_conv1}
\lim_{n\to \infty} |\E[u(\tilde{W}^n_n,\varphi^n_n,S^n_n)] - \E [u(W_1,\varphi_1,S_1)]| = 0
\end{eqnarray}
and
\begin{eqnarray}
\label{temp_conv2}
\lim_{n\to \infty} |\E[u(W^n_n,\varphi^n_n,S^n_n)] - \E [u(\ddot {W}^n_n,\varphi^n_n,S^n_n)]| = 0\,.
\end{eqnarray}
Our assertion is now proved by (\ref{temp_step_1_0}), (\ref{temp_conv1}), and (\ref {temp_conv2}). 
\end{proof}

\vspace{0.3cm}
\begin{proof}[Proof of Proposition \ref {thm1_pro2}]
We also assume $t=1$. Take any 
$(\zeta _r)_{0\leq r\leq 1} \in \mathcal {A}_1(\varphi )$ and define 
$(\psi ^n_l)_{l=0}^{n-1} \in \mathcal {A}_n^n(\varphi )$ by 
\begin{eqnarray*}
\psi ^n_l=\int_{(\frac{l-1}{n})\vee 0}^{\frac{l}{n}}\zeta _r dr. 
\end{eqnarray*}
Furthermore, we define $(\hat{\zeta} ^n_v)_v\in \mathcal {A}_1(\varphi )$ 
by \,$\hat{\zeta} ^n_v = 0\ \,\,(0\leq v \leq 1/n)$, 
$\zeta _{v-\frac{1}{n}}\ \,\,(v> 1/n)$. 
Let $(W^n_l,\varphi ^n_l,S^n_l)_{l=0}^{n-1} = \Xi ^n_n(w, \varphi , s ; (\psi ^n_l)_l)$, 
$(W_r,\varphi _r, S_r)_{0\leq r\leq 1} = \Xi _1(w, \varphi , s ; (\zeta _r)_r)$, and 
$(\hat{W}^n_r, \hat{\varphi}^n_r, \hat{S}^n_r)_{0\leq r\leq 1} $ $= \Xi _1(w, \varphi , s ; (\hat{\zeta}^n_r)_r)$.
We also let $X^n_l = \log S^n_l$, $X_r=\log S_r$ and $\hat{X}^n_r=\log \hat{S}^n_r$. 
Moreover, define $(\tilde{X}^n_r)_r$ by 
(\ref {def_of_tildeX}) replacing $(\hat{\psi }^n_l)_l$ with $(\psi ^n_l)_l$. 

Since $(\zeta _r)_r$ is left-continuous and 
bounded, 
we can apply Lebesgue's dominated convergence theorem to see that 
\begin{align*}
&\E\left[ \sup _{r\in [0,1]}
\left\vert \int ^r_0 
(g(\zeta_{v})- g(\hat{\zeta}^n_v))dL_v \right\vert 
\right]\leq 
\tilde{\gamma }\int ^1_0 \E \left[\left\vert g(\zeta_{v})-g(\hat{\zeta}^n_v) 
\right\vert \right] dv
\longrightarrow 0, \ \ n\to \infty\,. 
\end{align*}
Therefore, we can apply Lemma \ref{Ishi_Lem} with 
\begin{align*}
F^{n,1}_r=-\int_0^r g(\zeta _v) dv, \quad 
F^{n,2}_r=-\int_0^r g(\hat{\zeta}^n_v) dv
\end{align*}
and 
$\Pi _n=[0, 1]$ to obtain 
\begin{align}\label{IshiLem_result_3}
&\E \left[\sup _{r\in [0,1]} \vert X_r - \hat{X}^n_r \vert \right]
\longrightarrow 0 , \ \ 
n\rightarrow \infty . 
\end{align}
Using Lemma \ref{Lemma_Moment_Estimate}, (\ref{Inequality_Diff_ex_and_ey}), 
(\ref{IshiLem_result_3}), and Lebesgue's dominated convergence theorem, we have
\begin{align}\label{W_approx_1}
&\E [\vert W_1- \hat{W}^n_1 \vert ^{1/2}] \nonumber \\
& \quad
\leq \widehat{C}' \Big\{ \sqrt{\Phi_0} 
\Big( \E\Big[ \sup_{v\in [0,1]} \vert X_v - \hat{X}^n_v \vert \Big]\Big)^{1/2} 
+ \Big( \E\Big[\int_0^1 \vert \zeta_v-\hat{\zeta}^n_v \vert dv\Big]\Big)^{1/2} 
\Big\} \nonumber \\
&\quad \longrightarrow 0 , \ \ n\rightarrow \infty ,
\end{align}
where $\widehat {C}' = (3sC_{1, K})^{1/2}$. 

Next, 
let $F^{n,3}_r = - \sum _{l=0}^{\lceil nr - 1\rceil}g^n_l(\psi ^n_l)$ and 
$\Pi_n = \{ l/n ; l = 0,\cdots , n \}$. 
Then we have 
\begin{align}
\sup_{v\in \Pi_n(r)}| F^{n, 2}_v - F^{n, 3}_v |
\leq & 
\sum_{l=0}^{n-1} g_n(\psi ^n_l) \left | n L_{(l+1)/n}-n L_{l/n}-\tilde{c}^n_l \right | 
\nonumber \\
& + \sum_{l=0}^{n-1}\left |\frac{1}{n}g(n \psi ^n_l) - g_n(\psi ^n_l) \right| n (L_{(l+1)/n}-L_{l/n}) 
\label{Prop1_2_condition_for_IshiLem_1} \\
& + \int_{0}^{1} \vert g(\hat{\zeta}^n_v) - g(n \psi ^n_{[nv]}) \vert dL_v
\nonumber \\
& + g (\|\zeta \|_{\infty}) 1_{[0, 1]\setminus \Pi _n}(r) (L_{\lceil nr \rceil/n}-L_r ) , 
\quad r \in [0, 1].
\nonumber 
\end{align}
Then we see that 
\begin{align}\label{temp_H}
&\E \left[ \int_{0}^{1} \left\vert g(\hat{\zeta}^n_v) - g(n \psi ^n_{[nv]}) \right\vert dL_v \right]
\nonumber \\
&\quad \leq \tilde{\gamma } h(\|\zeta \|_{\infty}) 
\int_{0}^{1-\frac{1}{n}} 
\E \left[ \left\vert H_n (v) \right\vert \right] dv
\longrightarrow 0, \ \ n\to \infty ,
\end{align}
where 
\begin{eqnarray*}
H_n (v) = n \int_{[nv]/n}^{([nv]+1)/n} \zeta_u du - \zeta_v .
\end{eqnarray*}
By (\ref{Prop1_2_condition_for_IshiLem_1}), (\ref {temp_H}), 
and an argument similar to the proof of Proposition \ref{thm1_pro1}, 
we obtain 
\begin{align*}
\E \left [\sup _{v\in \Pi_n} | F^{n, 2}_v - F^{n, 3}_v |\right ] + 
\int _0^1 \E \left [\sup _{v\in \Pi_n(r)} | F^{n, 2}_v - F^{n, 3}_v |\right ] dr \longrightarrow 0 , \ \ 
n\rightarrow \infty . 
\end{align*}
Thus we get 
\begin{eqnarray}\label{IshiLem_result_4}
\E \left [\max _{l = 0, \ldots , n}|X^n_l - \hat{X}^n_{l/n}|\right ] =
\E \left [\sup _{v\in \Pi_n}| \tilde{X}^n_v - \hat{X}^n_v |\right ]\longrightarrow 0 , \ \ 
n\rightarrow \infty 
\end{eqnarray}
by virtue of Lemma \ref {Ishi_Lem}.

Define 
\begin{align*}
\ddot{W}_n^n &= w + n \sum_{l=0}^{n-2}\psi ^n_{l+1} 
\int_{l/n}^{(l+1)/n} \exp (\hat{X}^n_{v+\frac{1}{n}})dv . 
\end{align*}
Using (\ref{Inequality_Diff_ex_and_ey}), 
we get 
\begin{align}
&\E [\vert \hat{W}^n_1-\ddot{W}_n^n \vert ^{1/2}]
\leq \widehat{C} '
\left\{ \int _0^{1-\frac{1}{n}} \E [ \left\vert H_n (v) \right\vert ] dv \right\}^{1/2}
\longrightarrow 0 , \ \ n\rightarrow \infty . \label{W_approx_2} 
\end{align}
Moreover, 
using Lemma \ref{Lemma_Moment_Uniform_Continuity} and (\ref{IshiLem_result_4}), we have
\begin{align}
&\E [\vert \ddot{W}_n^n - W_n^n \vert ^{1/2}] 
\leq \widehat{C}'
\E \left[ \sum_{l=0}^{n-1} \psi ^n_l n \int _{l/n}^{(l+1)/n} 
\vert \hat{X}^n_v - X^n_l + g^n_l (\psi ^n_l) \vert dv \right]^{1/2} 
\nonumber \\ \nonumber 
&\quad \leq \widehat{C}' \Big\{ 
\Phi_0 \Big[ \Phi_0 \varepsilon _n \Big( \frac{C^*}{n}+\tilde{\gamma} \Big)
+\frac{g(\|\zeta \|_{\infty})}{n} 
+ \tilde{\gamma} h(\|\zeta \|_{\infty}) \int _0^{1-\frac{1}{n}} \E [ \left\vert H_n (v) \right\vert ] dv
\Big] \\\nonumber 
&\qquad \qquad 
+ \Phi_0 \Big[ \frac{\widetilde{C}_{2, K} }{n^{1/4} }
+\E \big [\max _{l = 0, \ldots , n}|X^n_l - \hat{X}^n_{l/n}|\big ] \Big] 
+ \frac{\tilde{\gamma} \|\zeta \|_{\infty} g(\|\zeta \|_{\infty}) }{n}
\Big\}^{1/2}\\
&\quad \longrightarrow 0 , \ \ n\rightarrow \infty . 
\label{W_approx_3} 
\end{align}
By (\ref{W_approx_1}), (\ref{W_approx_2}), and (\ref{W_approx_3}), we get 
$\lim_{n\to \infty} \E [\vert W_1 - W_n^n \vert ^{1/2}]=0$. 
Furthermore, using (\ref{IshiLem_result_3}) and (\ref{IshiLem_result_4}) 
we have \\$\lim_{n\to \infty}\E [\vert X_1 - X^n_n \vert]=0$ and 
$\lim_{n\to \infty}\E [\vert S_1 - S^n_n \vert ^{1/2}]=0$. 
Moreover, obviously it holds that $\varphi^n_n=\varphi_1$. 
Then we can apply Lemma \ref{lemm_conti_u} to obtain 
\begin{align}\label{result_of_lemma_conti_u_prop5}
\lim_{n\to \infty} \vert \E [u(W_1, \varphi _1, S_1)] - \E[u(W^n_n,\varphi^n_n,S^n_n)] \vert =0 .
\end{align}
Our assertion is now proved by (\ref{result_of_lemma_conti_u_prop5}) and the following inequality: 
\begin{align*}
\E [u(W_1, \varphi_1, S_1)] \leq 
\vert \E [u(W_1, \varphi _1, S_1)] - \E[u(W^n_n,\varphi^n_n,S^n_n)] \vert 
+ V^n_n (w, \varphi , s; u). 
\end{align*}
\end{proof}

\section{Concluding Remarks}\label{section_conclusion}

In this paper, we generalized the framework in \cite {Kato} and 
studied an optimal execution problem with random MI. 
We defined the MI function as a product of an i.i.d.\ positive random variable 
and a deterministic function in a discrete-time model. 
Furthermore, we derived the continuous-time model of an optimization problem as a limit of the discrete-time models, 
and found that the noise in MI in the continuous-time model can be described as a L\'evy process. 

We will investigate properties of the continuous-time value function in \cite {Ishitani-Kato_COSA2}.


\bibliographystyle{amsplain}

\end{document}